# Longitudinal forces in pinched electric currents

By F. O. Minotti

Departamento de Física, Facultad de Ciencias Exactas and Naturales, Universidad de Buenos Aires

and Instituto de Física del Plasma (CONICET), Buenos Aires, Argentina

*It is shown that the theory of Mbelek and Lachièze-Rey predicts longitudinal forces of gravitational origin in pinched current distributions, with magnitudes large enough to have noticeable effects.*

We consider a simple model of a uniform, axially symmetric current distribution along a straight wire of radius $a$, with total current $I$ flowing along the $z$ direction and radially limited to the region $0 \leq r \leq R(z)$, with $R(z) \leq a$. The corresponding (azimuthal) magnetic field is then:

$$B_\varphi(r,z) = \frac{\mu_0 I r}{2\pi R^2(z)}, \qquad r \leq R(z)$$

$$B_\varphi(r,z) = \frac{\mu_0 I}{2\pi r}, \qquad r \geq R(z).$$

According to the scalar-tensor theory of Mbelek and Lachièze-Rey (MLR) [1] this magnetic field gives rise to a gravitational potential whose equation is [2,3]:

$$\nabla^2 \chi = \kappa B_\varphi^2,$$

with $\kappa \approx 10^9$ (SI units), and where $\chi$ and its first derivatives are continuous at $r = R(z)$. It is to be noted that this equation applies even in the region with zero current because the correct boundary conditions for the scalar $\phi$ of MLR theory can only be satisfied by the non-zero laplacian solution of its determining equation (see related discussion in [2,3].)

For simplicity we consider a given pinch geometry of the form $R(z) = R_0 \exp(\lambda z)$, so that the corresponding equation for $r \leq R(z)$ is explicitly written as

$$\frac{1}{r}\frac{\partial}{\partial r}\left(r\frac{\partial \chi}{\partial r}\right) + \frac{\partial^2 \chi}{\partial z^2} = \kappa A r^2 \exp(-4\lambda z),$$

with

$$A = \frac{\mu_0^2 I^2}{4\pi^2 R_0^4}.$$

The complete solution of this equation, with the reference value of $0$ at $r = 0$, is easily obtained to be:

$$\chi_{int}(r,z) = \frac{\kappa A}{64\lambda^4}[J_0(4\lambda r) + 4\lambda^2 r^2 - 1]\exp(-4\lambda z), \qquad (1)$$

where $J_0$ is the Bessel function of order zero, and the subscript "int" refers to "internal" (the radial region occupied by the current). To more simply obtain the matching solution for $r > R(z)$ (the "external" solution) we consider the case of an elongated pinch with axial extension larger than its radial dimensions, so that $\lambda^2 a^2 \ll 1$. In this case the solution (1) can be approximated, Taylor developing the Bessel function, by

$$\chi_{int}(r,z) = \frac{\kappa A}{16} r^4 \exp(-4\lambda z). \qquad (2)$$

The "external" solution for $r > R(z)$ satisfies the equation

$$\frac{1}{r}\frac{\partial}{\partial r}\left(r\frac{\partial \chi}{\partial r}\right) + \frac{\partial^2 \chi}{\partial z^2} = \kappa \tilde{A} r^{-2},$$

with

$$\tilde{A} = \frac{\mu_0^2 I^2}{4\pi^2}.$$

The solution of this equation with the correct continuity conditions with the "internal" solution (2) at $r = R(z)$, is then readily seen to be

$$\chi_{ext}(r,z) = \frac{\kappa A}{16} R_0^4 + \frac{\kappa A}{4} R_0^4 \ln\left[\frac{r}{R(z)}\right] + \frac{\kappa \tilde{A}}{2} \ln^2\left[\frac{r}{R(z)}\right], \qquad (3)$$

where the condition $\lambda^2 a^2 \ll 1$ was used (the second order $z$ derivative turns out to be much smaller than the radial ones).

The axial force on the axial extension $\Delta z$ of the pinch is finally given by

$$F_z = -\rho \int_0^{\Delta z} dz \int_0^{R(z)} 2\pi r \frac{\partial \chi_{int}}{\partial z} dr - \rho \int_0^{\Delta z} dz \int_{R(z)}^{a} 2\pi r \frac{\partial \chi_{ext}}{\partial z} dr,$$

where $\rho$ is the wire mass density. The direct evaluation of this expression, using the solutions (2) and (3), gives the result

$$F_z = \frac{\rho \kappa \mu_0^2 I^2 a^2}{48\pi}\left\{11 + 3\ln(\gamma) - \frac{24\ln(\gamma)}{\gamma} - \frac{11}{\gamma^2}\right\}, \qquad (4)$$

where $\gamma \equiv \frac{a}{R_0}$, and where it was used that, according to the chosen pinch profile, the extension $\Delta z$ of the pinch satisfies $\lambda \Delta z = \ln(\gamma)$. In terms of surface stresses $\sigma$ one can write, from (4),

$$\sigma = \frac{\rho \kappa \mu_0^2 I^2}{48\pi^2}\left\{11 + 3\ln(\gamma) - \frac{24\ln(\gamma)}{\gamma} - \frac{11}{\gamma^2}\right\}.$$

For Graneau's experiment [4] one has $\rho = 2.7 \times 10^3 \, kg/m^3$, with a peak current of about $7 \, kA$. With these values, a pinch with $\gamma = 5$ is enough to reproduce the surface stress of $4 \, N/mm^2$ calculated by Johansson (in his MS thesis [5]) for this experiment, while a $\gamma = 100$ generates a stress three times larger, which could possibly break a thermally stressed wire.